\documentclass[a4paper,11pt]{article}
\pdfoutput=1 % if your are submitting a pdflatex (i.e. if you have
\usepackage{jheppub} % for details on the use of the package, please
\usepackage[T1]{fontenc} % if needed
\usepackage{amssymb,amsmath,amsbsy}
\usepackage{mathrsfs,verbatim,enumerate}
\usepackage{graphicx}
\usepackage[utf8]{inputenc}

\title{\boldmath Time-Delayed Electrons from Higgs Decays to Right-Handed Neutrinos}

\author{John D. Mason}

\affiliation{Western State Colorado University\\ 1 Western Way, Gunnison, CO 81231}

\emailAdd{jmason@western.edu}

\abstract{We consider the physics reach at the high luminosity LHC (HL-LHC) using the timing capability of a minimum ionizing particle (MIP) timing layer and the electromagnetic calorimeter in the CMS experiment for a simplified right-handed neutrino effective theory. In the simplified model, the lightest right-handed neutrinos are produced through higgs decay with mean decay lengths of order $\mathcal{O}(0.1~{\rm m}-100~{\rm m}$) and then subsequently decay into final states that include electrons. We consider the effect of several different kinematic cuts for a timing based analysis of this model. We demonstrate that if timing information can be successfully incorporated into the event trigger, it is possible to probe the higgs branching ratio to two right-handed neutrinos to the order $\mathcal{O}(10^{-5}) $ for a range of different right-handed neutrino lifetimes. }

\begin{document} 
\maketitle
\flushbottom

\section{Introduction and Overview}

The origin of neutrino mass remains an un-determined aspect of the Standard Model.  It is well-known that a type-I seesaw mechanism \cite{Minkowski:1977sc} \cite{Mohapatra:1979ia} can efficiently generate the hierarchically small neutrino masses ($m_{\nu} \leq 0.1 ~{\rm eV}$ \cite{Tanabashi:2018oca}) in a theory that is renormalizable up to the electroweak scale and beyond (see \cite{Mohapatra:2006gs} for a review). In models where the right-handed neutrinos have an electroweak scale majorana mass\cite{Pilaftsis:1991ug,Belotsky:2002ym,Han:2006ip,Hung:2006ap}, it is natural for the right-handed neutrinos to have mean decay lengths of order $\mathcal{O}(0.1~{\rm m} -100~{\rm m})$ due to small coupling to standard model particles \cite{Graesser:2007yj}. Since the right-handed neutrinos in such models typically decay to final states with leptons, recent investigations of such scenarios have focussed on the displaced-vertex/track nature of the final state leptons \cite{Graesser:2007yj,Graesser:2007pc,Helo:2013esa,Maiezza:2015lza,Accomando:2016rpc,Caputo:2017pit,Deppisch:2018eth,Cottin:2018kmq,Cottin:2018nms,Drewes:2019fou,Liu:2019ayx} and have proven to be an efficient probe of this class of models.

In general, if long-lived particles produce a displaced-vertex/track type signature at colliders, they can also be associated with calorimeter energy deposition that is ``time-delayed" with respect to the energy deposition of massless states produced at the primary collision vertex. Such a time-based signature is not new but in fact a classic signature of low-scale gauge mediated supersymmetry breaking (GMSB) \cite{Dine:1994vc} and the focus of a number of experimental searches \cite{Toback:2004xd,Abulencia:2007ut,Aaltonen:2008dm,CMS-PAS-EXO-12-035}. The proposal to include a minimum ionizing particle (MIP) detector on the inner barrel and endcap regions of the electromagnetic calorimeter (Ecal) as well as planned improvements to the timing resolution of the Ecal itself at the CMS experiment promises to increase time resolution for charged particles to 30 ps at the HL-LHC \cite{cms_mip}.  The utility of this MIP/Ecal timing capability has been recently analyzed in \cite{Liu:2018wte} to investigate the reach of the timing capability to GMSB and higgs decays to long-lived particles where peak sensitivity is achieved at $c\tau$ = 1 m. Since a timing based analysis need not depend crucially on tracking, the time-delayed signature can arise from a signal event so long as it occurs in the main inner detector volume. Contrast this with the displaced-vertex/track searches which typically reach peak sensitivity at $c\tau $=  1-10 cm for final state electrons. The distinction is due to the deterioration of tracking capability as electron final states produced from long-lived particles that decay farther than a distance of 50 cm transverse to the beamline \cite{Chatrchyan:2014fea}. Additionally, the volume of the region of efficient tracking is roughly only 10\% of the volume enclosed by the inner detector. This opens up the possibility that a timing based analysis may provide a complimentary probe of models having displaced-vertex/tracking signatures, in particular targeting models with longer lifetimes where a majority of the delayed decays occur outside of the region where tracking is most efficient. 

The purpose of this paper is to analyze the physics reach of a timing based analysis of right-handed neutrino models where the higgs boson can decay to ``long-lived" right-handed neutrinos which subsequently decay to final states with electrons. We shall work within the context of the type I seesaw effective theory described in \cite{Graesser:2007yj,Graesser:2007pc} using the timing analysis techniques of \cite{Liu:2018wte}.  This model is a well-motivated member of a much larger class of models recently reviewed in \cite{Alimena:2019zri}, in which the higgs decays to long-lived particles. In order to fully utilize MIP/Ecal timing capability we will not rely on track reconstruction of the electrons in the signal. It will be critical for our purposes that the MIP/Ecal timing capability be incorporated into the event trigger at the CMS experiment and we shall assume for the remainder of the paper that this will be realized at the HL-LHC. Since the performance of such a timing trigger is not finalized, we will need to make reasonable assumptions regarding the efficiency of such a trigger, and we shall clearly state them below. The paper is organized as follows. In section \ref{theory} we describe the electroweak scale seesaw effective theory and then in section \ref{simple} focus on a particular region of parameter space where only one right-handed neutrino is phenomenologically relevant, parameterizing a simplified model in terms of three parameters. In section \ref{signal}, we describe the general kinematic features of new higgs decay channels that are present in this model.  Based on these features, in section \ref{sim} we outline search criteria, describe our event simulations, estimate backgrounds, and present results in terms of exclusions on higgs branching ratios to right-handed neutrinos as a function of lifetime for different masses. We conclude in section \ref{Conclusion}.

\section{Electroweak Scale Type I Seesaw Effective Theory }\label{theory}

\begin{figure}[tbp]
\centering
\includegraphics[scale=0.7]{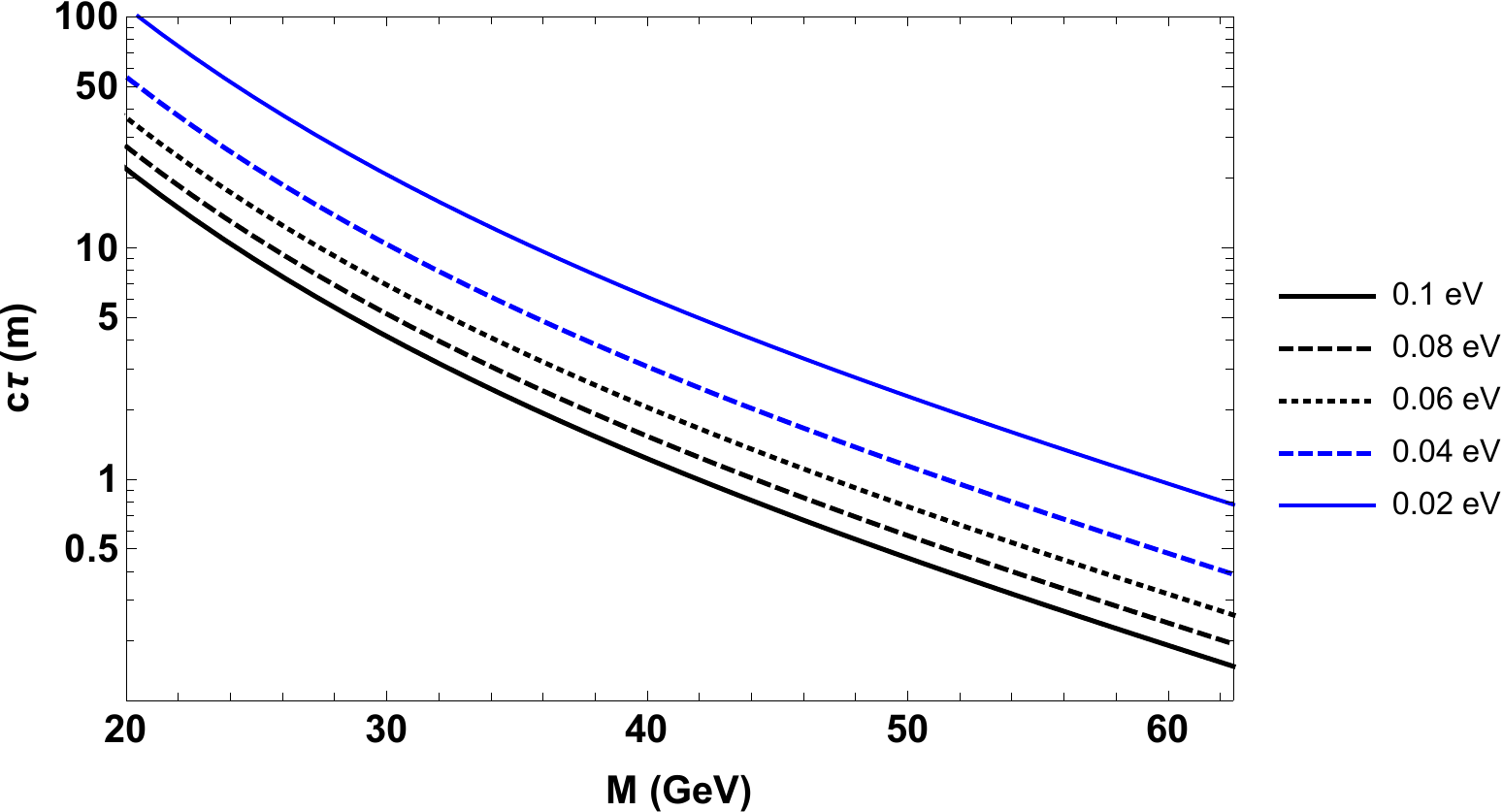} 
\caption{ Right-handed neutrino mean decay length (m) as a function of right-handed neutrino Majorana mass (GeV) for different fixed values of physical neutrino mass }
\label{ctau1}
\end{figure}

The field content of the model is the standard model plus three majorana ``right-handed" neutrinos:  $N^I$, where I = 1, 2, 3  labels the three right-handed neutrino flavors. In addition to the particle content, the model also has the following renormalizable interactions: 
\begin{equation} \label{Ln}
-\mathcal{L}_N = (y)_{iJ}\left( \tilde{H} \cdot \overline{L_L}^i \right) N_R^J + \frac{1}{2}M_{IJ}\left( \overline{N_L^c}^I N_R^J \right) + \rm{h.c.}
\end{equation}
In this notation $\tilde{H} = i \sigma_2 H^*$ where $H$ transforms under $SU(2)_L \times U(1)_Y$  as a $({\bf 2}, +1)$, gauge indices are not shown and $i = 1, 2,3 ~(I,J = 1, 2, 3 )$ are light (heavy) neutrino flavor indices and the $R,L$ subscripts correspond to Right and Left projections (e.g. $N^I_{R,L} = P_{R,L}N^I$). It is well known that at energies below $M_{IJ}$, these 
interactions generate a dimension-five Weinberg operator and yield Majorana neutrino masses, $m_{\nu}$, given by
\begin{equation} 
m_{\nu} \sim \frac{(yv)^2}{M_N}, 
\end{equation}
where $v = 174~{\rm GeV}$ is the vacuum expectation value of the higgs field and $M_N$ ($y$) is the scale of the majorana mass matrix $M_{IJ}$ ($y_{iJ}$). If $M_N$ is of order the electroweak-scale, neutrino masses of order $\mathcal{O}(0.1~\rm{eV})$ will be generated for yukawa couplings $y \sim 5\times 10^{-7}$. Due to the mass mixing induced in the full neutrino sector of this theory, the neutral and charged current interactions include W and Z interactions with the right handed-neutrinos: 
\begin{equation} \label{LWZ}
-\mathcal{L}_{W/Z} \supset \frac{g}{\sqrt{2}}W_{\mu}^+({\bf V})_{Ji}\left( \overline{\tilde{N^c}}^{J}\gamma^{\mu}P_L\tilde{l}^i \right)  +\frac{g}{2\cos{\theta_W}}Z_{\mu}({\bf \tilde{V}})_{iJ}\left( \overline{\tilde{\nu}}^i\gamma^{\mu}P_L\tilde{N}^{cJ} \right). 
\end{equation}
Here ${\bf V} $ and ${\bf \tilde{V} } $ are $3\times3$ matrices that depend on the $M_{IJ}$, $(y)_{iJ}$ and rotation matrices that diagonalize the neutrino and lepton sector and the tilde indicates that the field is associated with the mass eigenstate. The yukawa interaction of \eqref{Ln} gives rise to higgs mediated decays of the right-handed neutrinos:
\begin{equation} \label{Lh}
-\mathcal{L}_h\supset \frac{h}{v}{\bf V}^*_{iJ}M_J\overline{\tilde{N}^c}^J P_L \tilde{\nu}_i. 
\end{equation}
We briefly review the derivation of these terms and the precise definitions of ${\bf V} $ and ${\bf \tilde{V} } $ in Appendix A. Parametrically, the couplings  ${\bf V} $ and ${\bf \tilde{V} }$ are of order $\mathcal{O}(\frac{yv}{M})$. We shall focus on a right-handed neutrino mass, ${\rm 20 ~GeV} < M_{N_1} < 60 {\rm GeV}$, in which case the dimensionless couplings in  ${\bf V} $ and ${\bf \tilde{V} } $ are $\mathcal{O}( 10^{-6})$. The phenomenological effect of such a small coupling is to give the right-handed neutrino a relatively large lifetime of order $\mathcal{O}( 0.1-100~ {\rm ns})$. We graph the parametric relation between the scale of the left handed neutrino mass, the right handed neutrino mass and the mean decay length of the right-handed neutrino in Figure \ref{ctau1}.

Interestingly, there is considerable overlap of the parameter space for the right-handed neutrino mean decay length and the size of the CMS inner detector, where the shortest distance from the interaction point (IP) to the inner radius of the electromagnetic calorimeter (Ecal) is $1.29$ meters. Note also that significant portions of parameter space for this model have $c\tau > 1{\rm m}$. This opens up the opportunity to probe the nature of the right-handed neutrinos using precision timing. As the neutrino model presented above presently stands, all couplings are too small to give any appreciable cross-section for production of right-handed neutrinos. Of course in order to probe the nature of the right-handed neutrinos, there must be a mechanism for them to be produced. We now turn to motivate and parameterize such a mechanism. 

The model so far includes renormalizable interactions. However, if we view the above theory as an effective theory, we should include higher dimensional operators suppressed by a mass scale $\Lambda$ where this effective theory breaks down. If $\Lambda$ is not too much greater than the electroweak scale, then these higher dimension operators can affect phenomenology. Presumably this would be an adequate description of some neutrino models that extend the most basic type-I seesaw with other particles. However, matching existing models to this effective theory is beyond the scope of this paper. Ignoring flavor indices, there are only two linearly independent gauge invariant dimension-five operators involving the right-handed neutrino, $N_R^I$ \cite{Graesser:2007yj}
\begin{equation}
 ~~~~~\mathcal{O}^{IJ}_h = \frac{\left( H^{\dagger}\cdot H \right) \overline{N_L^c}^{I}N^J_R}{\Lambda} ~~~~~~\mathcal{O}^{IJ}_B = \frac{\overline{N_R^c}^{I} \sigma^{\mu \nu}N_R^J B_{\mu \nu}}{\Lambda},
\end{equation}
where $B_{\mu \nu}$ is the $U(1)_Y$ field strength. In this paper we shall restrict our attention to a regime of parameter space where the effects of $\mathcal{O}^{IJ}_B$ are negligible. Specifically, we shall work with only one phenomenologically accessible lightest right-handed majorana neutrino where $\overline{N^c}_{1R} \sigma^{\mu \nu}N_{1R}=0$ due to the majorana nature of the right-handed neutrino. Turning our attention to $\mathcal{O}^{IJ}_h$, one main phenomenological consequence results from including it in our description of the effective theory; when the higgs field gets a vev, $\mathcal{O}^{IJ}_h$ can lead to a significant production mode of right-handed neutrinos through higgs boson decay if the mass of the right-handed neutrino mass eigenstate is less than $m_h/2$. Approximately, 
\begin{equation}
BR(h \rightarrow N_1 N_1) \sim \frac{2v^4c^2}{3m_b^2\Lambda^2} \beta^3, 
\end{equation}
where $\mathcal{L} \supset c_{IJ}\mathcal{O}^{IJ}_h$, with $c_{11}=c$ and $\beta = (1-\frac{4 M^2_{N_1}}{m^2_h})^{1/2}$.  

If we take $\Lambda = 10 ~{\rm TeV}$ then $BR(h \rightarrow N_1 N_1)< 0.035$ for $c< 0.68$. The higgs vev also leads to a contribution to the right-handed neutrino mass, however that contribution will always be sub-dominant compared to the already present $M$-term in \eqref{Ln} for the parameters studied in this paper. 

To summarize, the completely general right-handed neutrino effective theory lagrangian considered in this paper is give by: 
\begin{equation} \label{lfull}
\mathcal{L} = \mathcal{L}_{SM} + \mathcal{L}_N + c_{IJ}\mathcal{O}^{IJ}_h,
\end{equation}
where $\mathcal{L}_{SM}$ is the standard model lagrangian which includes $\mathcal{L}_{W/Z}$ and $\mathcal{L}_h$ when all fields are rotated to their mass eigenbasis and $\mathcal{L}_N$ from \ref{Ln}. The parameter space of this model is quite large, depending on $y_{iJ}$, $M_{IJ}$ and $c_{IJ}$. In the next section we will restrict the parameter space of this model with some basic simplifying assumptions in order to more easily facilitate a collider study of the model.

\section{A Simplified Model }\label{simple}

We shall now restrict our attention to a particular region of parameters of the right-handed neutrino effective theory that is of phenomenological interest. The region of interest is chosen as follows: 
\begin{enumerate}
\item We assume the mass of the lightest right-handed neutrino ($\tilde{N}^1$), $M_{\tilde{N}^1} < ~62.5 ~{\rm GeV}$ and the other right-handed neutrinos have masses above 62.5 GeV and are thus inaccessible or otherwise produced with such low rates that they are not phenomenologically relevant. 
\item We assume that the lightest right-handed neutrino ($\tilde{N}^1$) couples dominantly to $e^{\pm}$ or $\nu_e$. We will impose this by making ${\bf V}_{11} = {\bf \tilde{V}}_{11} = \mathcal{O}(10^{-6})$ and all other ${\bf V}_{iJ} = {\bf \tilde{V}}_{iJ} = 0 $. Our choice for the value of this coupling comes from the requirement that the neutrino mass is given by the relation:  $m_{\nu} \sim (yv)^2/M_N$. 
\item Even with the above constraints, there remains enough freedom in the choice of $y_{iJ}$ and $M_{IJ}$  that the neutrino mass mixings can be reproduced without spoiling the above choices for couplings.
\end{enumerate}
The above restrictions applied to \ref{lfull} result in the following interaction lagrangian between the lightest kinematically accessible right-handed neutrino and standard model leptons, gauge bosons, and the higgs:
\begin{equation}
\begin{aligned}
\mathcal{L}_{int} &= -\frac{g}{\sqrt{2}}W_{\mu}^+({\bf V})_{11}\left( \overline{N_1^c}\gamma^{\mu}P_L\tilde{e} \right)  -\frac{g}{2\cos{\theta_W}}Z_{\mu}({\bf V})_{11}\left( \overline{\nu}_e\gamma^{\mu}P_LN_1^{c} \right) \\ &-\frac{h}{v}{\bf V}^*_{11}M_1\overline{N^c_1} P_L \nu_e+\frac{2cv}{\Lambda}h \overline{N_1^c}N_1 - \frac{1}{2}M_{N_1}\overline{N_1^c}N_1 +{\rm h.c.}
\label{lint}
\end{aligned}
\end{equation}
where we have now lowered the flavor index on the right-handed neutrino and removed the ``L,R" labels as well and the ``$\sim$" for neatness. The lagrangian \eqref{lint} depends effectively on only three parameters $({\bf V_{11}}, M_{N_1}, \frac{cv}{\Lambda} )$. However, if the see-saw mechanism is responsible for neutrino masses, then there is a constraint between two of the parameters in terms of the light neutrino masses ($m_{\nu}$):  ${\bf V_{11}}^2 = \frac{m_{\nu}}{M_{N_1}} \sim 10^{-11}$ for $M_{N_1} \sim 10 ~{\rm GeV}$ and $m_{\nu} \sim 0.1 ~{\rm eV}$. Restricting our attention to masses in the range: $20~{\rm GeV} < M_{N_1} < 62.5 ~{\rm GeV}$, this constraint combined with the above interactions will lead to three-body decays of the right-handed neutrinos through off-shell W/Z/h to a number of final states with decay lengths ($c\tau$) that range between: $0.1 ~{\rm m} < c\tau < 100~{\rm m} $ (see Figure \ref{ctau1}). Note that due to range of $m_{\nu}$ that may be physically realized in nature, a number of different values of ${\bf V_{11}}$ ( and thus $c\tau$ ) are possible given a choice of mass $M_{N_1}$.  The third parameter, $\frac{cv}{\Lambda}$, controls the branching ratio of the higgs boson to a pair of right-handed neutrinos. For $\frac{cv}{\Lambda}$ in the range $10^{-4} <\frac{cv}{\Lambda}<10^{-2} $, the branching ratio, $BR(h \rightarrow N_1 N_1)$, ranges between $ 2.5 \times 10^{-6} < BR(h \rightarrow N_1 N_1) < 2.5 \times 10^{-2}$. Since this acts as an additional decay mode for the higgs, the indirect bound on $h \rightarrow {\rm invisible }$ \cite{Aad:2015gba} constrains the $BR(h \rightarrow N_1 N_1)< 0.27$ in this model.  
We shall exchange the parameters that appear in the lagrangian for the corresponding phenomenological parameters: 
\begin{equation}
\left( {\bf V_{11}}, M_{N_1}, \frac{cv}{\Lambda} \right) \rightarrow \left( c\tau, M_{N_1}, BR \right)
\end{equation}
where $BR = BR(h\rightarrow N_1 N_1)$ and describe this simplified model in terms of the phenomenological parameters for the remainder of this work. 

\begin{figure}
\centering
\includegraphics[scale=0.20]{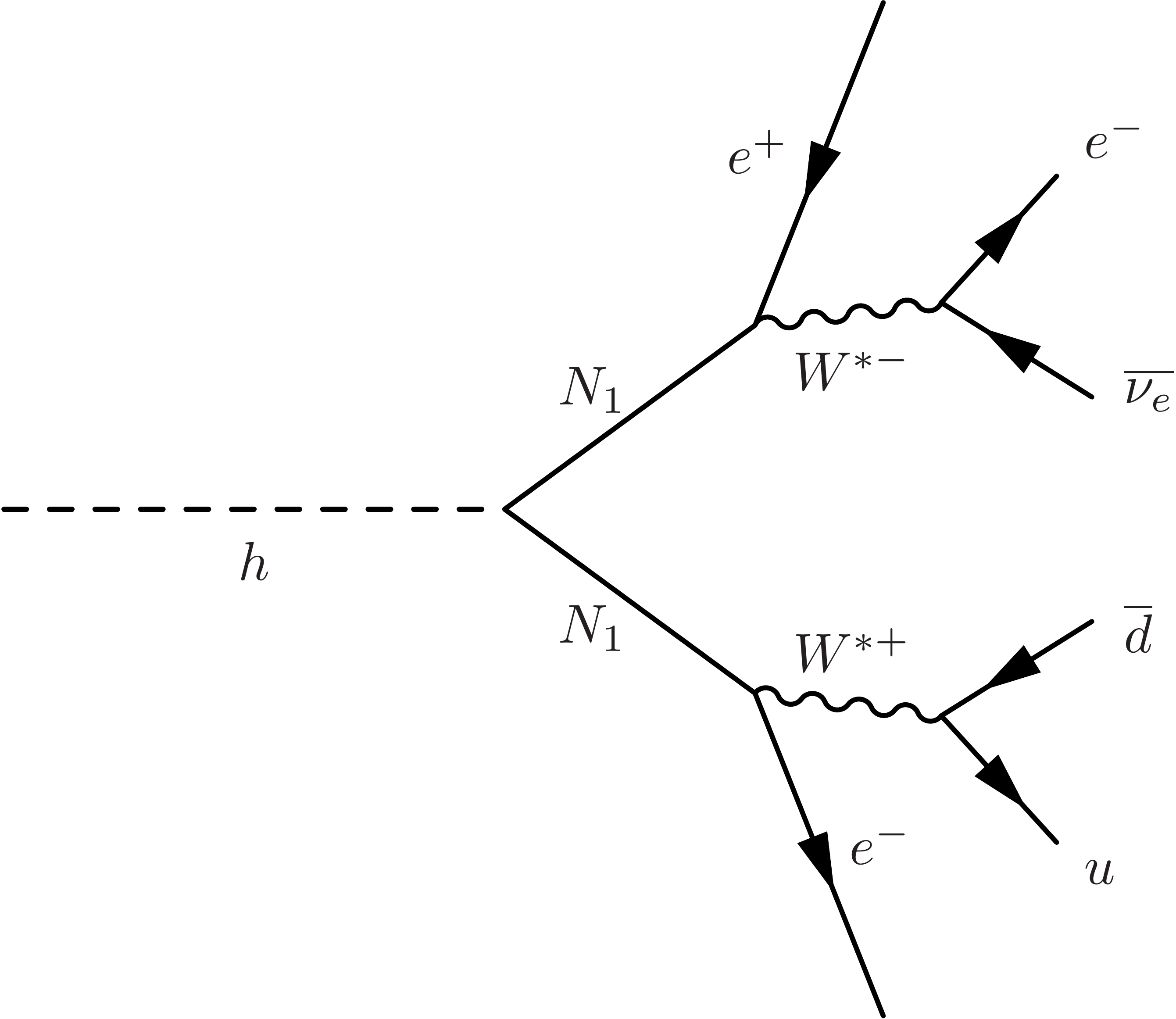} 
\caption{ An example of a higgs decay process through off-shell W-bosons into final states with electrons.}
\label{feyn}
\end{figure}

The only appreciable production mode open to the right-handed neutrinos is through the intermediate production and decay of the higgs boson. After being produced in pairs by some fraction of higgs decays, the lightest right-handed neutrinos have a variety of decay modes open to them. For the assumptions listed above, the decays will always be three-body decays through off-shell Z, W or h. A sample decay is shown in Figure \ref{feyn}. Roughly $75\%$ of the time there will be at least one $e^{\pm}$ and jets or missing energy associated with the final state. See \cite{Graesser:2007pc} \cite{Duarte:2016miz} \cite{Duarte:2015iba} for a detailed description of final states. 

\section{General Signal Properties}\label{signal}

\begin{figure}
\centering
\includegraphics[scale=0.50]{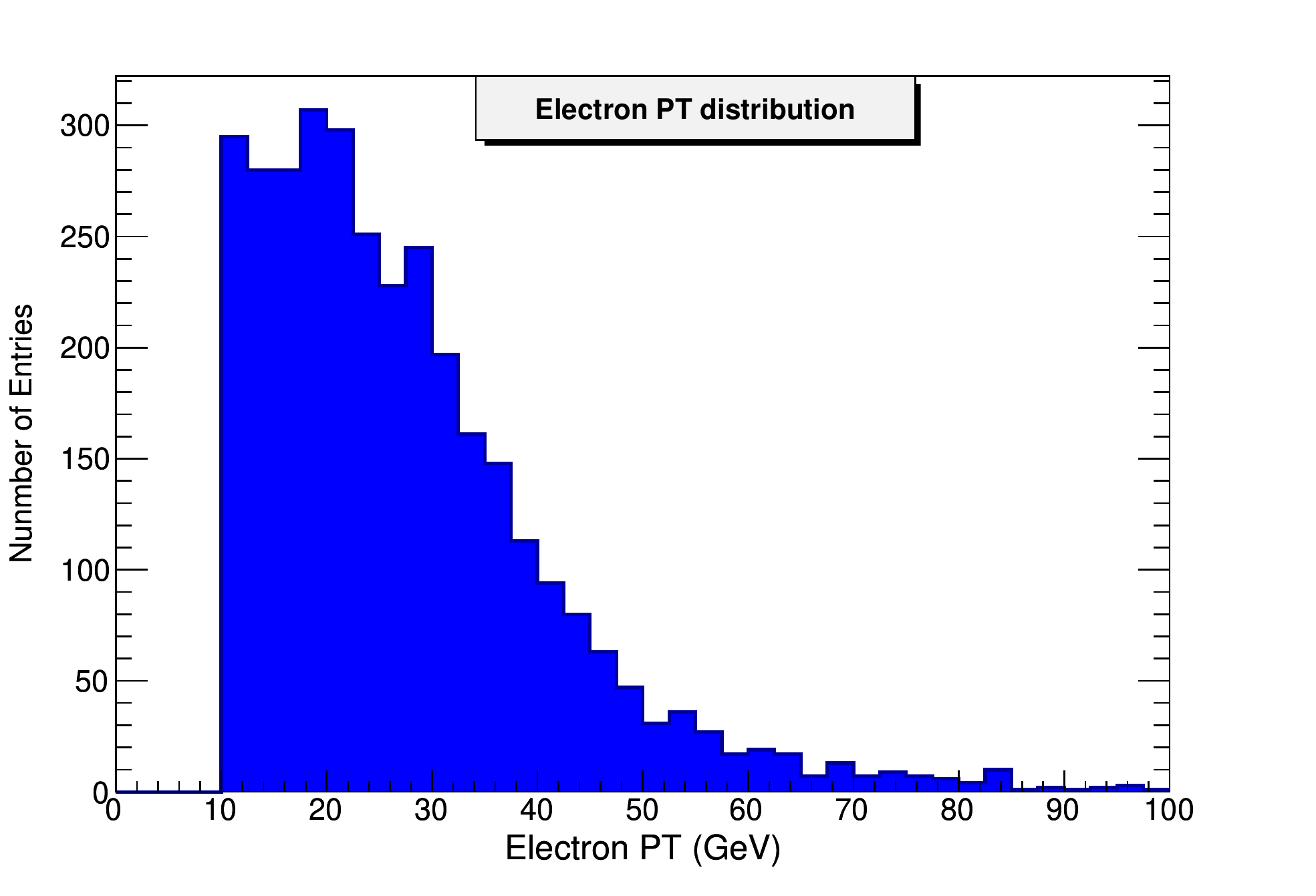} 
\caption{ $p_T$ distribution of the leading electron from right-handed neutrino decays ( $M_N = 50~{\rm GeV}$). In this case, the default Delphes CMS card efficiencies have been applied. These include, for instance, zero percent efficiency for electron $p_T$s below 10 GeV. }
\label{ept}
\end{figure}

Signal events produced in the simplified model typically possess electrons, jets and missing energy. The most important of these for our purposes are the electrons. The electrons produced from right-handed neutrino decays typically have a relatively small transverse momentum as can be seen in Figure \ref{ept}, falling off quite rapidly for $p_T >  40 ~{\rm GeV}$. 
The origin of this feature is the fact that the higgs boson only has 125 GeV of rest energy to give its multiple decay products, resulting in the soft leptons and jets. This leads to a low efficiency for such events to pass the existing Level 1 + High Level Trigger at ATLAS \cite{atlas_trigger} or CMS \cite{Khachatryan:2016bia}.

Another important feature of these decay modes is the displaced character of the right-handed neutrino decays. The charged electron tracks will not originate at the primary interaction point. The efficiency for track reconstruction falls quickly as the true location of the origin of the electron track becomes greater than a distance of $R=50~{\rm cm}$ transverse to the beamline \cite{Chatrchyan:2014fea} \cite{CMS:2014hka}. Thus a significant fraction of the decay events will produce either poorly re-constructed electrons or events that resemble photons in that the electrons deposit energy in the Ecal but are not associated with tracks after being passed through the multi-level track reconstruction algorithm. These issues are carefully considered in searches for models that produce leptons with tracks having large transverse impact parameters \cite{Khachatryan:2014mea} \cite{CMS:2014hka}.  

\begin{figure} 
\centering
\includegraphics[scale=0.50]{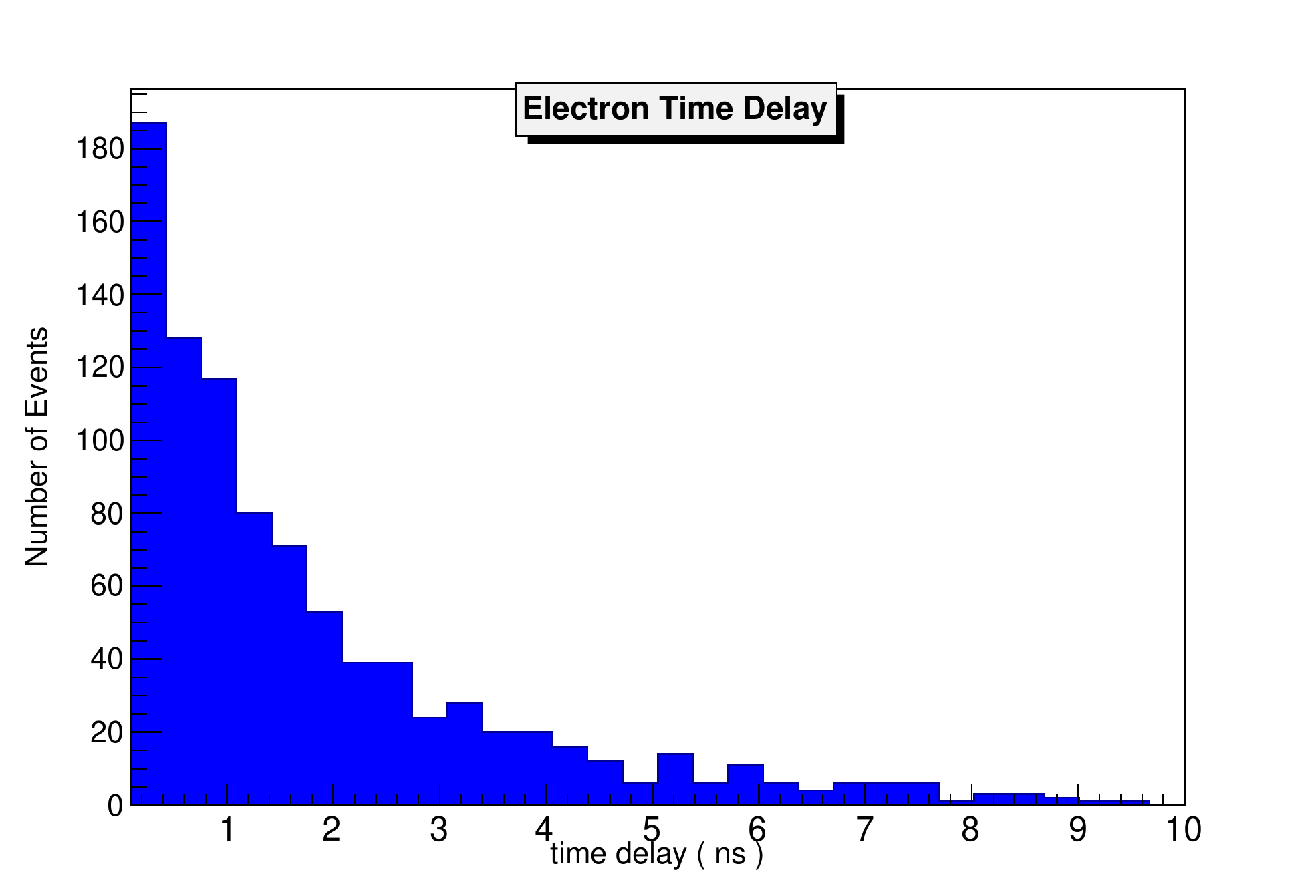} 
\caption{ Distribution of the time delay,  $\Delta t = t_{arrival} - t_{prompt}$, of an electron originating from the decay of a $50~{\rm GeV }$ right-handed neutrino with a mean decay length of $c\tau=1.0~{\rm m}$ . }
\label{etime}
\end{figure}

Closely related to the displaced nature of the decays, and most important for our purposes, is the ``delayed" time at which electrons from right-handed neutrino decays deposit energy in the Ecal. For instance, after being produced, the right-handed neutrinos will travel some time $t_N$ before decaying into electrons and jets. The electrons then travel an additional distance, $L$, from their production vertex location to an Ecal cell located at some $(\eta, \phi)$, in a time, $t_e = L/c$. Thus the arrival time is given by: $t_{arrival} = t_N + t_e$.  Had the electrons originated promptly from the primary vertex, they would have a prompt arrival time of $t_{prompt} = \frac{1.29m}{2c\sin(\tan^{-1}(e^{-\eta}))}$ for an Ecal cell located at $(\eta, \phi)$ within the central barrel.  With these parameters, a time delay:
\begin{equation}
\label{tdel}
 \Delta t = t_{arrival} - t_{prompt}
 \end{equation}
 can be associated with each Ecal energy deposit in an event. A distribution of $\Delta t$ is shown in Figure \ref{etime} for signal events of a 50 GeV right-handed neutrino with a 1 meter mean decay length. For heavier mass right-handed neutrinos, which tend to be produced with lower velocities, the time delay is largely due to the longer lifetime of the particle. For lighter mass right-handed neutrinos, which are produced with higher boosts, the time delay is largely due to the longer ``kinked"  path taken by the right-handed neutrino and subsequently produced electron. The right-handed neutrino decays we are discussing here are very similar to the classic signature for low-scale gauge mediated supersymmetry breaking models \cite{Dine:1994vc}, although the soft kinematics are certainly unique to the delayed neutrino scenario. 

\section{ Search Proposal Using MIP Timing at CMS }\label{sim}

We now propose a strategy to search for events predicted by the simplified model described in section \ref{simple}:  higgs decays to two ``long-lived" right-handed neutrinos which in turn decay to electrons, jets and neutrinos through interactions with charged and neutral currents. Previous work on this model has focused on the displaced nature of the electron and muon tracks \cite{Caputo:2017pit}. Here we focus on the ``delayed" nature of the events and follow very closely the analysis described in \cite{Liu:2018wte}. We first define our search criteria, describe our event simulation then we estimate backgrounds and give out results.

\subsection{Search Criteria} 
We now consider ways that the HL-LHC can capitalize on the unique kinematics of delayed right-handed neutrino decays and be sensitive to these models using precision timing information. A key component of our analysis will be the proposed MIP timing layer \cite{cms_mip} and Ecal timing resolution for which the final design, calibration and installation is anticipated at the HL-LHC. The Ecal timing resolution is projected to measure the time of energy deposition at the inner radius of the Ecal with a 30 ps time resolution when $p_T \geq 50 ~{\rm GeV}$. The MIP timing layer will have similar timing resolution for charged particles with much smaller $p_T$. The timing layer's main purpose is to veto events that are out of time with the hard interaction as a means to reduce pile-up. However, the spread of pile-up from one bunch crossing leads to a 190 ps time spread for all of the energy deposition in the Ecal for that crossing. With 25 ns between bunch crossings, this leaves roughly 12.5 ns of time to remain sensitive to Ecal energy deposition before the detector must be prepared for the next bunch crossing. This opens up the possibility for the MIP/Ecal timing information to be incorporated into the trigger such that if a event shows significant out of time energy deposition, greater than 1 ns for instance, then such an event can be triggered on rather than vetoed. Furthermore, we shall include objects identified as a photon \emph{or} as an $e^{\pm}$ in our analysis since we do not want counting of signal events to rely solely on objects that are associated with well reconstructed tracks. In the following, we shall assume a timing trigger will be implemented and let the current level-1 trigger menu \cite{Khachatryan:2016bia} act as a guide for how we should model the timing trigger performance and efficiency in combination with the $p_T$ of the object. We will model the timing trigger as 50\% efficient for events with out of time Ecal energy deposit greater than 1 ns ( $\Delta t > 1 ~{\rm ns}$ and $p_T$ roughly equal to the level-1 trigger threshold). This will be our standard timing trigger benchmark. However, since the timing trigger's performance has not been finalized, we shall vary the trigger threshold around this benchmark value to demonstrate the search's sensitivity to different trigger thresholds. In order to register a reference time for each event, we require a jet produced from initial state radiation with $p_T> ~30~ {\rm GeV}$. Based on the location in $\eta$ of this jet, its time of arrival, and the location of the primary vertex an expected time of arrival for any promptly produced massless particle, $t_{prompt}$, can be assigned to any Ecal energy cell given it's value of $\eta$.  Given the above considerations, we define our benchmark search criteria as: 
{\bf Event Selection Criteria ($e^{\pm}$ or $\gamma$)}
\begin{enumerate}
\item A leading $e^{\pm}$ or $\gamma$ with: $p_T > 20 ~{\rm GeV}$, $|\eta| < 2.5$ and  $\Delta t > 1 ~{\rm ns}$  
\item At least one ISR jet with $p_T > 30 ~{\rm GeV}$
\end{enumerate}
These criteria also meet or exceed our assumptions regarding the trigger threshold as discussed above.      

\subsection{Event Simulation} 
The effective right-handed neutrino model is implemented using FeynRules \cite{Alloul:2013bka} based on a modification of the model file built in \cite{Alva:2014gxa}\cite{Degrande:2016aje} and used for  \cite{Atre:2009rg}. Events are generated using MadGraph \cite{Alwall:2014hca} and Pythia 8.2 \cite{Sjostrand:2014zea}. Detector simulation is handled with Delphes \cite{deFavereau:2013fsa} using the CMS detector card. Events are produced in higgs production through gg fusion from pp collisions at 13 TeV center of mass energy with a scaled production cross-section of $44~{\rm pb}$. 

\begin{figure}
\centering
\includegraphics[scale=0.7]{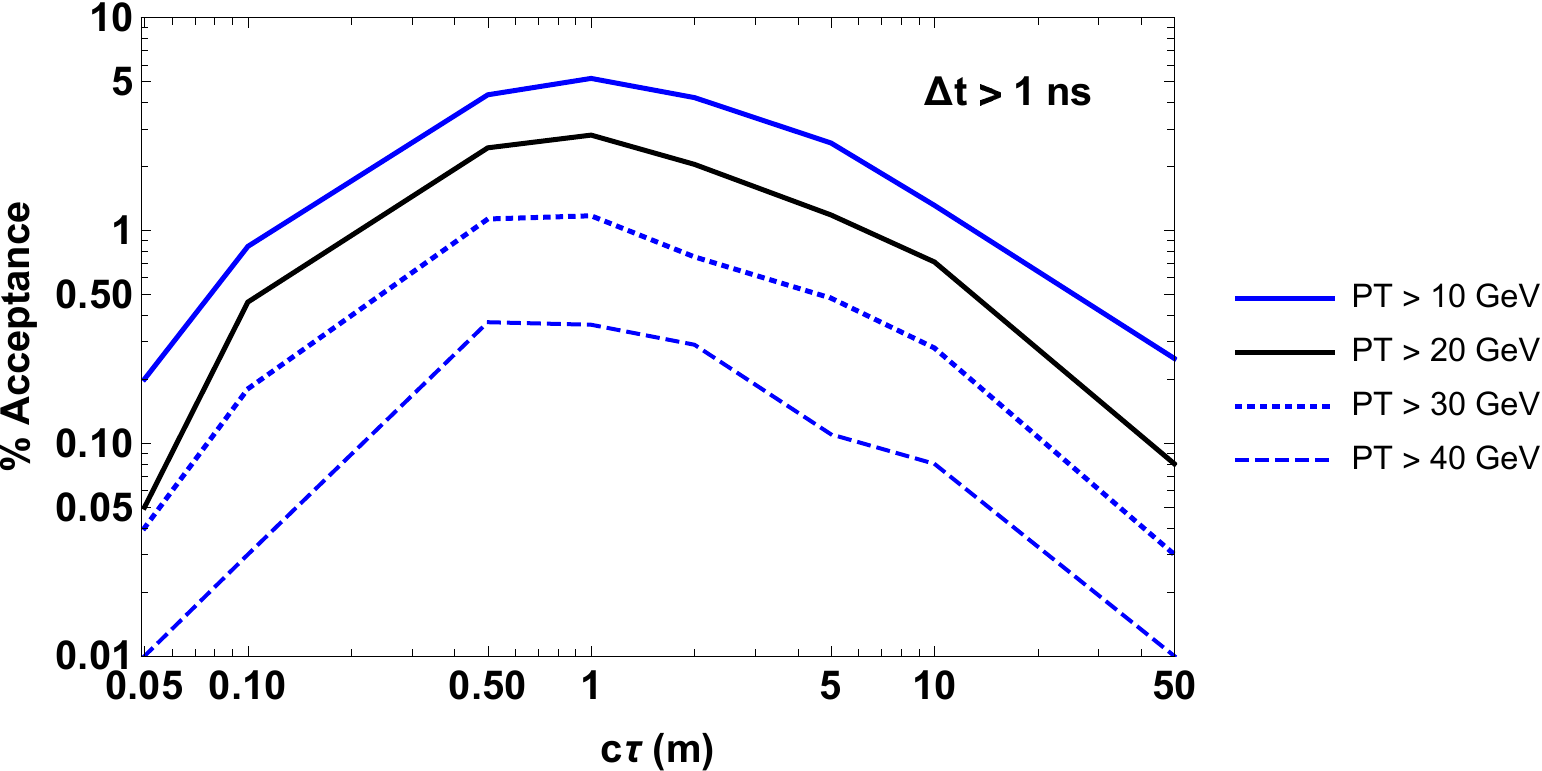} ~~~~~~~~
\includegraphics[scale=0.7]{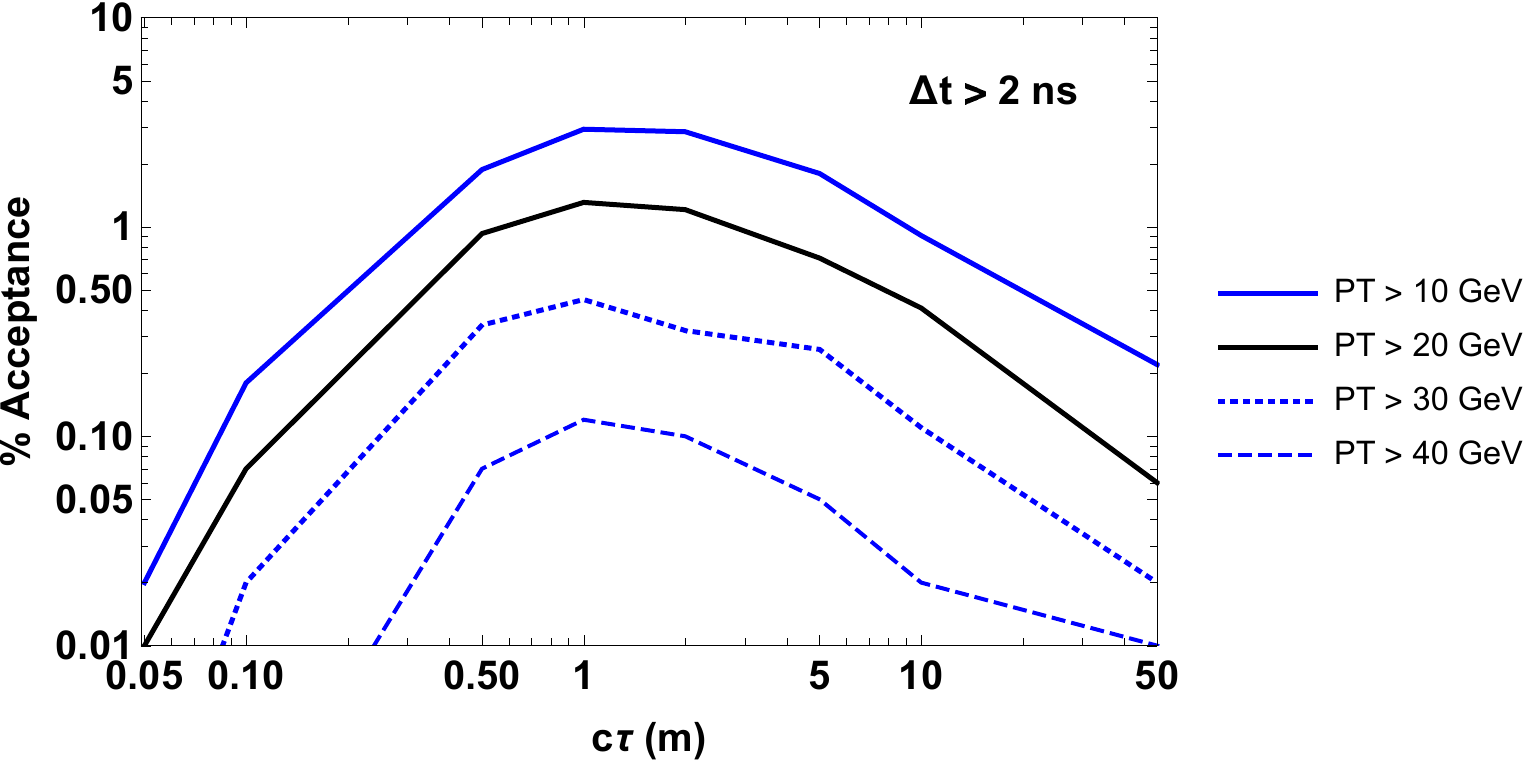} 
\caption{ Acceptance as a function of average decay length (m) for a right-handed neutrino mass, $M = 50 ~ \rm{GeV}$ for two different timing cuts ($\Delta t > 1~{\rm ns}$ and $\Delta t > 2~{\rm ns}$) in an Ecal timing based analysis. }
\label{acceg}
\end{figure}

We will define signal acceptance as the percentage of signal events that pass the cuts listed in the previous section \emph{after} being run through the standard Delphes detector simulation using the default CMS card. Due to $\eta$ and $p_T$-dependent efficiencies on electrons and photons, detector efficiencies are also incorporated into our signal acceptance values.  The signal acceptance for the search criteria is plotted as the darker of the lines in the left plot of Figure \ref{acceg}. This left figure also shows the acceptance as the $p_T$ trigger threshold is varied and the right shows the same variations for a $\Delta t$ threshold of 2 ns. As expected, the acceptance decreases significantly as the $p_T$ threshold is increased. Increasing the $P_T$ threshold from 20 GeV to 30 GeV (40 GeV) decreases the acceptance by a factor of three (ten).  
Including tracking information to identify the electrons results in an order of magnitude less acceptance for the same kinematic cuts. The reason for this is the requirement that the right-handed neutrino decays within $R< 50 ~{\rm cm }$ restricts the decays to fall within only 10\% of the volume of the inner detector significantly reducing sensitivity to model parameter space with long lifetimes. Furthermore, large time delays of $\Delta t > 1 ~{\rm ns} $ are less likely when the right-handed neutrinos decay so close to the interaction point. In these cases, we expect that an order of magnitude fewer signal events would pass cuts that utilize a timing based analysis in combination with a displaced-vertex/track type of analysis since they are somewhat mutually exclusive in this way an do not consider a tracking based analysis in this work. 

\begin{figure}
\centering
\includegraphics[scale=0.7]{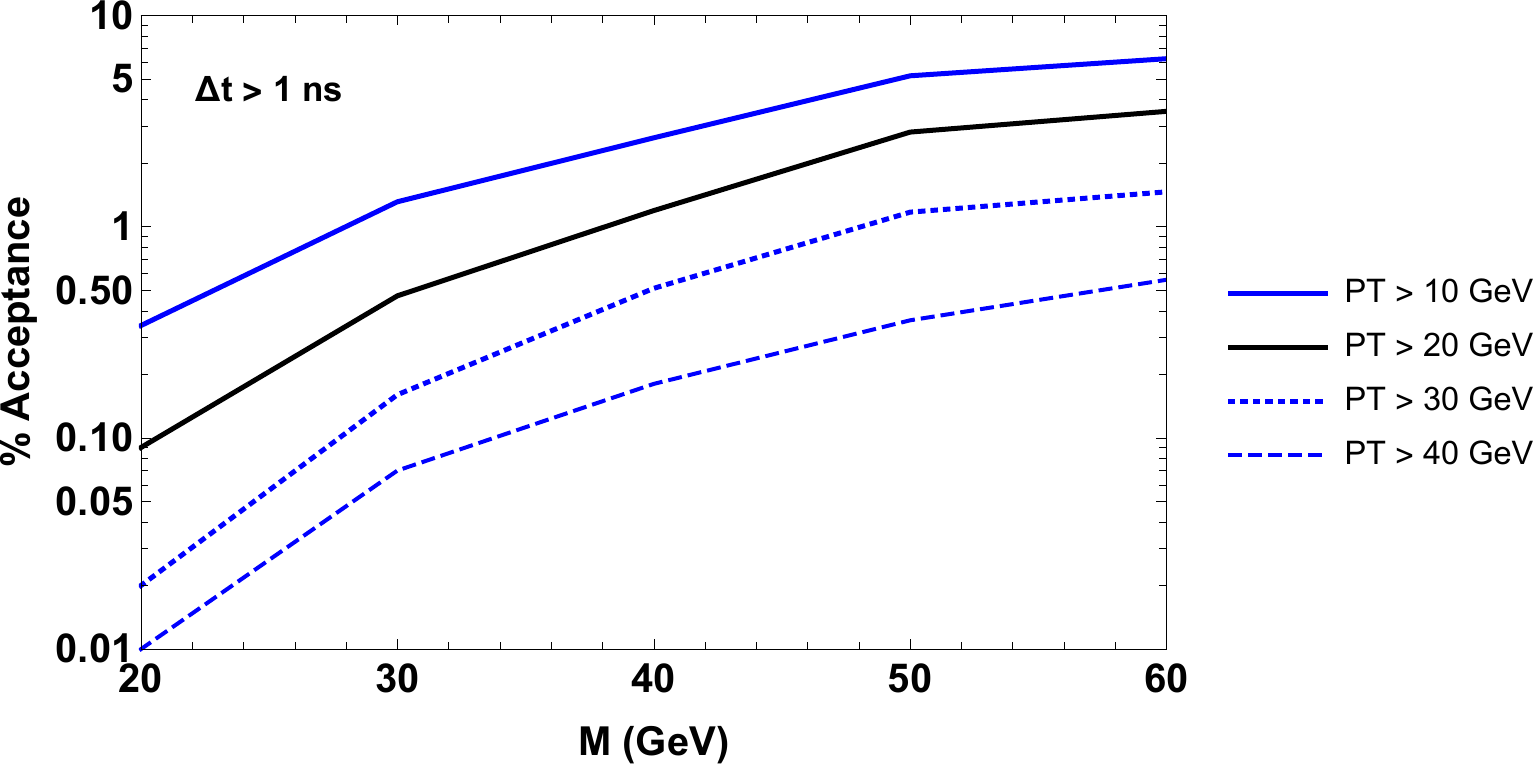}
\caption{ Acceptance as a function of right-handed neutrino mass for a decay length, $c\tau = 1 ~ \rm{m}$ for Ecal timing based analysis. }
\label{accm}
\end{figure}

Figure \ref{accm} shows the signal acceptance as a function of right-handed neutrino mass for fixed mean decay length ($c\tau = 1~{\rm m}$) where acceptance is larger for heavier masses. Lighter right-handed neutrino masses are produced with larger boosts from higgs decay compared to heavier right-handed neutrinos. This results in less time delay due to the geometric similarity of this decay process to that of promptly produced particles. 

Generating events at leading order in Madgraph and then scaling up the cross-section with using the appropriate k-factor, as we have done above, overpopulates (under-populates) the higgs' low $p_T$ (higher but less than 100 GeV $p_T$) distribution. Explicitly including the $gg \rightarrow h+j$ process in an NLO simulation would have two main results; it would harden the $p_T$ distributions above and reduce the ``kinked" path contribution to time delay due to the additional boost of provided to the higgs. We expect that these effects would somewhat compensate each other in the acceptances above resulting in a change of no more than $10\%$ in our results \cite{deFlorian:2016spz}. 

\subsection{ Backgrounds }\label{bckg}

Ordinary Standard Model processes in pp collisions are generically prompt compared to nano-second time scales and do not produce events with significant values for the time delay ($ \Delta t$). However, jet and photon production cross-sections are so large for $p_T \sim 20 ~{\rm GeV}$ that very unlikely timing mis-measurements can be realized.  A main source of background is time of arrival mis-measurement due to timing resolution and pile-up due to other pp collisions that occur during the same bunch crossing. We shall model these backgrounds with  gaussian distributions closely following \cite{Liu:2018wte} and review and apply their estimates with slight modifications below. 
To estimate background arising from jets we utilize the low-$p_T$ data of \cite{Khachatryan:2016mlc} which gives a jet cross-section of $\sigma_j = 1.6 \times 10^{8}~{\rm pb}$ for $p_T > 30~{\rm GeV}$.  The cross-section for inclusive photon production is $\sigma_{\gamma} = 47 \times 10^3 ~{\rm pb}$ \cite{Aad:2016xcr}.  Other important values for estimating backgrounds rates are the fraction of jets that are track-less, $f_j = 10^{-3}$, the rate for a jet to fake a photon, $f_{\gamma} = 10^{-4}$ \cite{jetfake}, and a full integrated luminosity of $\mathcal{L}= 3 \times 10^{6} ~{\rm pb}^{-1}$. Production of a photon and jet can happen through actual photon production in association with a jet or through dijet production where one of the jets is mis-identified as a photon; we estimate the numbers of such events as $N_{\gamma} = \sigma_{\gamma} \mathcal{L} $ and $N_{j} = \sigma_j \mathcal{L} f_{\gamma}$, respectively. Our estimate is $N_{bg} = \sigma_{\gamma} \mathcal{L} +\sigma_j \mathcal{L} f_{\gamma} = 2\times10^{11}$. In fact this is conservative due to the fact that photon cross-section used in inclusive and would be less if a $p_T$ cut was put on an associated jet. If we model a $\delta = 30 ~{\rm ps}$ time resolution with a gaussian smear ( $\frac{dP}{d\tau} = \frac{1}{\sqrt{2 \pi}\delta}e^{-\tau^2/{2\delta^2}}$), the number of background events estimated to be mis-measured as having a time delay of greater than 1 ns is negligible since 1 ns is more than 30 standard deviations outside this distribution. Anticipating less resolution in timing for lower $p_T$ objects since the $p_T$ of signal electrons are only $20~{\rm GeV}$, we note that even a 60 ps resolution in timing would result in negligible background.  A more relevant background however arises from pile-up. A signal-like event can arise when in addition to the collision that occurs to trigger the event, there is another collision in the same bunch crossing that does not produce tracks and remains unidentified. In this case, the trackless jets associated with this second collision can be interpreted as out of time energy deposition in the ECAL. An estimate of the number of these events can be given by: $N^{(PU)}_{j} = \sigma_j \mathcal{L} (n \sigma_j / \sigma_{inc} f_{\gamma} f_j)$ where $\sigma_{inc} = 80~{\rm mb}$. The factor of  $f_{\gamma}$ is the rate for a jet to fake a photon and the other factor of $f_j$ is the rate for the recoiling jet to be trackless. This results in $N^{(PU)}_{j} = 10^7$. Applying a gaussian smear of 190 ps resolution to model the timing distribution, where the value of 190 ps is determined by the longitudinal spread of the bunch crossing, the background estimate for $\Delta t > 1~{\rm ns}$ ($\Delta t > 2~{\rm ns}$) is 0.7 (zero) from pileup. 

Additional backgrounds that can lead to out of time signatures are from beam halo effects and satellite bunch crossings. Substantial suppression of these backgrounds was achieved in \cite{CMS-PAS-EXO-19-001}. We will assume that by utilizing similar techniques for this search these backgrounds could be similarly suppressed.

\subsection{Results}

\begin{figure}
\centering
\includegraphics[scale=0.8]{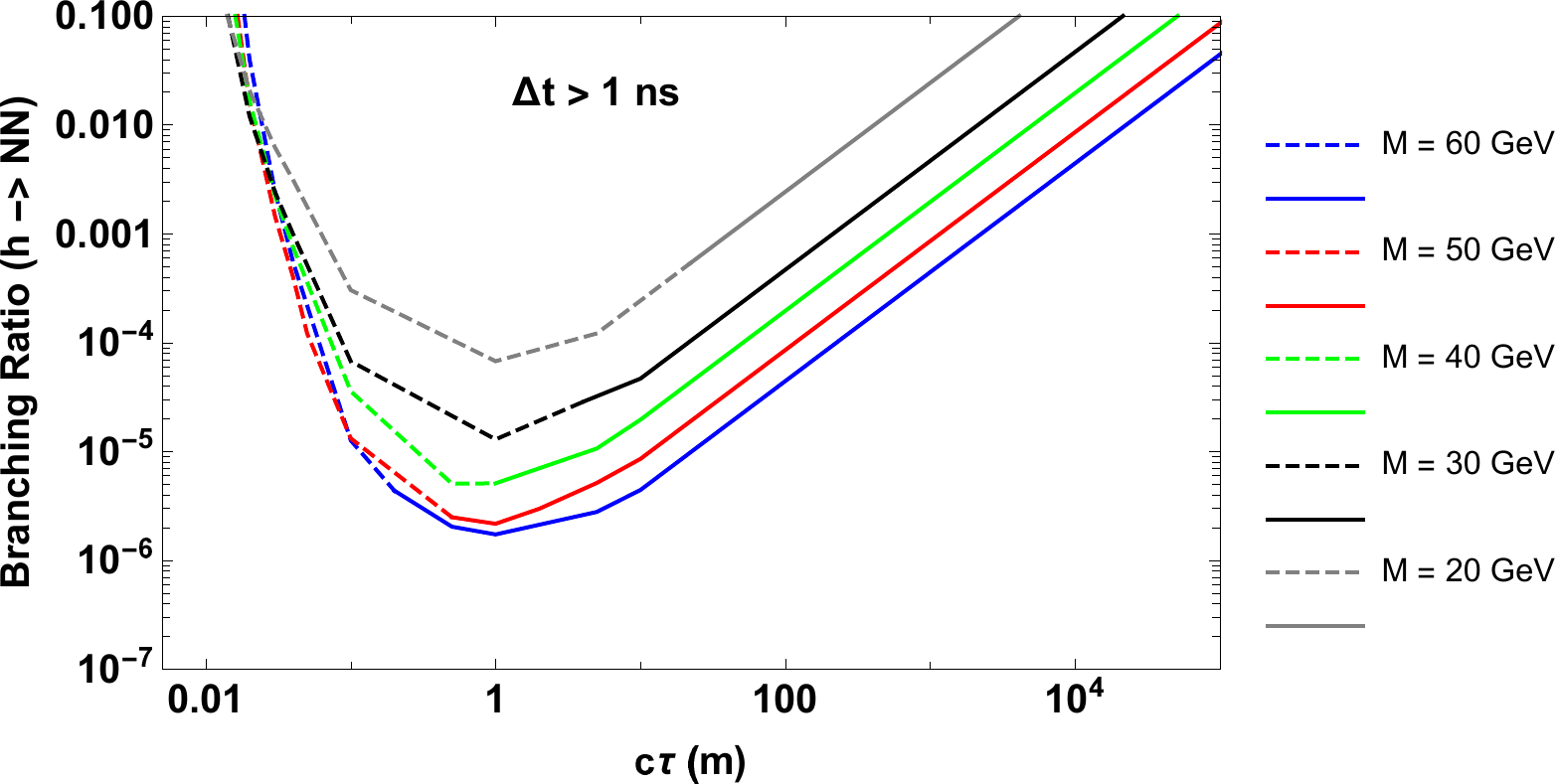}
\includegraphics[scale=0.8]{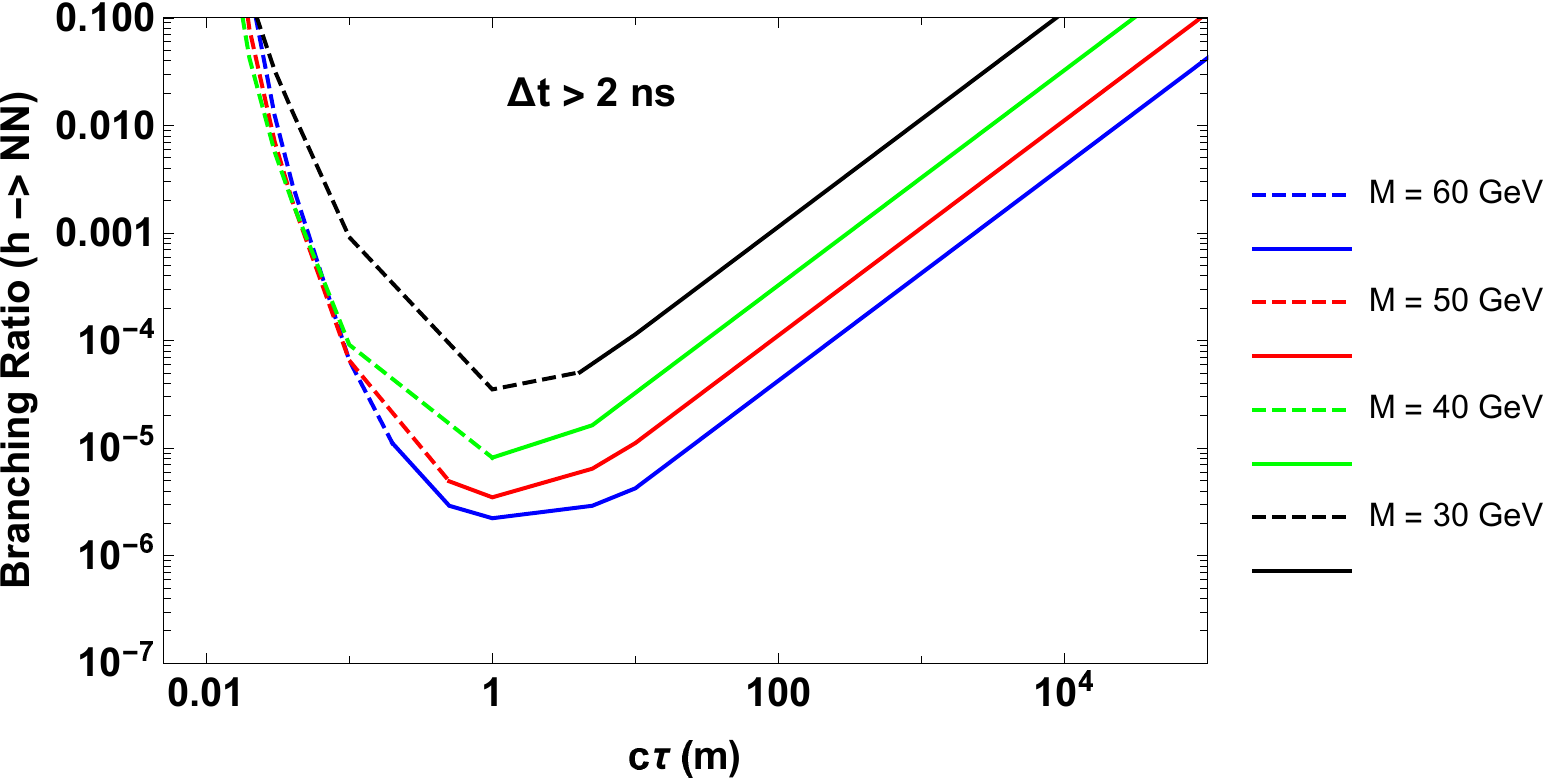}
\caption{ 95\% C.L. exclusion limits of higgs branching ratio to two right-handed neutrinos as a function of $c\tau$ for different right-handed neutrino masses in the simplified model. The transition from Dashed to Solid lines marks the $c\tau$ values at which the physical neutrino mass arising from the see-saw mechanism is $m_{\nu} \sim 0.1~{\rm eV}$. The top plot is for $\Delta t > 1 ~{\rm ns}$; the bottom plot is for $\Delta t > 2 ~{\rm ns}$ } 
\label{excl}
\end{figure}

Using the signal acceptances discussed above, we can characterize the experimental reach of the HL-LHC in terms of an exclusion bound on the Higgs branching ratio to two right-handed neutrinos for a given right-handed neutrino mass and lifetime. The 95\% C.L. exclusion is defined for models that predict at least four (three)  events at the 13 TeV HL-LHC with the full integrated luminosity of 3 $ab^{-1}$ for a timing cut of $\Delta t > 1~{\rm ns}$ ($\Delta t > 2~{\rm ns}$); this corresponds roughly to a 95\% C.L. exclusion where the $\Delta t > 1~{\rm ns}$ ($\Delta t > 2~{\rm ns}$) has a background of 0.7 (zero). In what follows we will consider the case with a timing trigger for events with $p_T > 20~{\rm GeV}$ and $\Delta t > 1 ~{\rm ns}$ and assume a flat trigger efficiency of $\epsilon_{trig} = 0.5$. In Figure \ref{excl} we plot the value of the higgs Branching Ratio ($BR_{hN_1N_1}$) that produces an average of four events that pass the kinematic cuts:  $p_T > 20~{\rm GeV}$ , $|\eta|< 2.5$, $\Delta t > 1 ~{\rm ns}$ in the top plot and  $p_T > 20~{\rm GeV}$ , $|\eta|< 2.5$, $\Delta t > 2 ~{\rm ns}$ in the bottom plot at the HL-LHC at 13 TeV with and integrated luminosity of $3 ~{\rm ab}^{-1}$. Since the physical neutrino mass parametrically depends on the higgs-lepton-neutrino yukawa coupling as $m_{\nu}\sim (yv)^2/M_N$, then as the right-handed neutrino lifetime is decreased for a given right-handed neutrino mass, the physical neutrino mass increases. At some lifetime for each mass, this parametric dependence dictates that $m_{\nu} \sim 0.1~{\rm eV}$. This is indicated in each plot at the transition from solid to dashed lines. While this bound is somewhat model dependent,  physical neutrino masses much larger than $0.1 ~{\rm eV}$ are disfavored by standard cosmology \cite{Aghanim:2018eyx}. As expected, timing sensitivity peaks near  $c\tau = 1 ~{\rm m} $. For either the $\Delta t > 1 ~{\rm ns}$  or $\Delta t > 2 ~{\rm ns}$ cuts, the sensitivity is best for a higgs branching ratio of $BR_{hN_1N_1} = 2\times10^{-6}$. The slightly lower background of the $\Delta t > 2 ~{\rm ns}$ cut compensates the lower acceptance, resulting in similar limits. This demonstrates that the signal is generally robust against moderate increases of the timing cut. Looking back at the discussion around Figure \ref{acceg}, one can expect that the sensitivity will decrease by a factor of three (ten) as the $p_T$ cut on the photons is increased to 30 GeV (40 GeV).  Finally we mention that for $c\tau$ in the range: $0.1~{\rm m} < c\tau < 10~{\rm m}$, exclusions were derived directly from simulation. Outside of this range however, we have extrapolated the bounds to reduce computational time.  We use a simple model of exponentially small likelihood of the number (N) of long decay lengths for $c\tau < 0.1~{\rm m}$: $\frac{dN}{dt} \sim e^{-t/(\gamma \tau)}$ and simple geometric dependence for the number (N) of short decay lengths: $ N \sim (c\tau)^{-1}$ when $c\tau > 10~{\rm m}$. The boost factor for lighter right-handed neutrinos leads to a more slowly decreasing exclusion for smaller mean decay lengths.  

A variation of the effective right-handed neutrino model discussed here was previously considered in \cite{Caputo:2017pit} where a displaced track analysis reached peak sensitivity at $c\tau \sim 0.1 ~{\rm m}$ and became weaker for longer lifetimes. That analysis showed a gap in sensitivity in the $1~{\rm m} < c\tau < 10~{\rm m} $ range before the MATHUSLA detector \cite{Curtin:2018mvb} sensitivity could potentially take over for larger mean decay lengths. This gap remains after scaling up the analysis of \cite{Caputo:2017pit} to the anticipated integrated luminosity at HL-LHC. Based on the sensitivities shown in Figure \ref{excl}, we have shown that a timing-based analysis is a promising avenue to fill that gap in sensitivity to better understand and constrain the right-handed effective neutrino model discussed in this work.

\section{Conclusion}\label{Conclusion}

We have shown that there exists an opportunity to probe a simplified right-handed neutrino effective theory by utilizing MIP/Ecal timing at the CMS experiment at HL-LHC. Generally, the final states of higgs decay can include soft electrons in the final state whose time of arrival at the MIP layer and Ecal is delayed with respect to the particles produced at the primary vertex. If MIP/Ecal timing can be incorporated in such a way as to trigger on events with greater than a nano-second time delay and with $p_T$ near the existing level-1 trigger threshold, then the CMS experiment can reach sensitivities of higgs branching ratios of order $\mathcal{O}(10^{-5})$ for a range of physically motivated right-handed neutrino masses and lifetimes for the simplified model described above.  For mean decay lengths of $c\tau  > {\rm 10 ~m}$, other experiments (e.g. the MATHUSLA detector \cite{Curtin:2018mvb}) are likely to provide better sensitivity to this model. For decay lengths of $c\tau  < {\rm 1 ~m}$ we expect displaced vertex/tracking would provide better sensitivity. However for the intermediate range $1~ {\rm ~m} < c\tau < {\rm 10~ m} $, we have shown that MIP/Ecal timing information can lead to considerable sensitivity to the type-I seesaw effective theory. Depending on the ability to incorporate timing into the event trigger, timing may provide the best limit for for this range of mean decay lengths.

\acknowledgments
I would like to thank Samuel Alipour-Fard, Matthew Citron, Nathaniel Craig, Seth Koren and Dave Sutherland for useful conversations regarding this work and for comments on the manuscript. I am especially indebted to Physics Department at UCSB for their hospitality in hosting me while on sabbatical from Western State Colorado University.

\appendix
\section{Right-Handed Neutrino Notation}

In this appendix we briefly describe the charged and neutral current interactions arising in the type-I seesaw model using the notation of \cite{Atre:2009rg}. The particle content is that of the Standard Model plus three right-handed neutrinos. The fermions in the lepton sector are: 
\begin{equation}
L^i = \left( \begin{array}{c} \nu^i_L \\ l^i_L \end{array}\right) ~~~~~(i = 1,2,3)~~~~~~ \rm{and}~~~~~~~~N^I_R~~~(I = 1,2,3)
\end{equation}
In this notation the charged and neutral current interactions are:
\begin{equation}
-\mathcal{L} \supset \frac{g}{\sqrt{2}}W_{\mu}^+\overline{\nu}_L^i\gamma^{\mu}l_L^i +\frac{g}{\cos{\theta_W}}Z_{\mu}\overline{\nu}_L^i\gamma^{\mu}\nu^i_L
\end{equation}
In addition to the other standard model interactions, the model has the following additional interactions at the renormalizable level:
\begin{equation}
-\mathcal{L}_N = (y)_{iJ}\left( \tilde{H} \cdot \overline{L_L}^i \right) N_R^J + \frac{1}{2}M_{IJ}\left( \overline{N_L^c}^I N_R^J \right) + \rm{h.c.}
\end{equation}
The higgs vev ( $v = 174~ \rm{GeV}$ ) yields mass terms for the neutrinos as follows: 
\begin{equation}
\begin{aligned}
-\mathcal{L}_{m} &= \frac{1}{2} \left( (yv)_{iJ} \overline{\nu_L}^iN^J_R + (yv)_{Ij} \overline{N^c_L}^I\nu^{cj}_R +M_{IJ}\overline{N_L^c}^IN_R^J \right) + \rm{h.c.} \\
& = \frac{1}{2} \left( \begin{array}{cc} \overline{\nu_L}^i & \overline{N_L^c}^{I} \end{array} \right) \left( \begin{array}{cc} 0 & (m_d)_{iJ} \\ (m_d)_{Ij} & M_{IJ} \end{array} \right) \left( \begin{array}{c} \nu^{cj}_R \\ N_R^J \end{array} \right) + \rm{h.c.}\\
& = \frac{1}{2} \left( \begin{array}{cc} \overline{\nu_L}^i & \overline{N_L^c}^{I} \end{array} \right) \mathcal{M} \left( \begin{array}{c} \nu_R^{cj} \\ N_R^J \end{array} \right)+ \rm{h.c.}
\end{aligned}
\end{equation}
with $(m_d)_{iJ} = v(y)_{iJ}$. The matrix $\mathcal{M}$ can be diagonalized using Takagi factorization as $\mathcal{M}_D = L^{\dagger}\mathcal{M}L^*$ and the mass eigenstates given by:
\begin{equation}
 \left( \begin{array}{c} \nu_L^{i} \\ N_L^{cI} \end{array} \right)= L\left( \begin{array}{c} \tilde{\nu}^{j} \\ \tilde{N}^{cJ} \end{array} \right),
\end{equation}
where $\tilde{\nu}$ and $\tilde{N}$ are mass eigenstates and L is given by: 
\begin{equation}
L = \left( \begin{array}{cc} U_{ij} & V_{iJ} \\ X_{Ij} & Y_{IJ} \end{array} \right). 
\end{equation}
As a result of this mixing the left handed neutrino interaction eigenstates shift slightly to include a small component of right-handed neutrino. Specifically, 
\begin{equation}
\Delta \nu^i_L = V_{iJ}\tilde{N}^{cJ}_{L} = V_{iJ}P_L\tilde{N}^{cJ}~~~~~~~~ {\rm and ~equivalently} ~~~~~~~~~~~~\Delta \nu^{ci}_R = V^*_{iJ}\tilde{N}^{J}_R=V^*_{iJ}P_R\tilde{N}^{J}.
\end{equation}
Writing the charged and neutral current interactions in terms of neutrino mass eigenstates then introduces the following interactions:
\begin{equation}
\mathcal{L} \supset \frac{g}{\sqrt{2}}W_{\mu}^+(V^{\dagger}O_L)_{Ji}\left( \overline{\tilde{N^c}}^{J}\gamma^{\mu}P_L\tilde{l}^i \right)  +\frac{g}{2\cos{\theta_W}}Z_{\mu}(U^{\dagger}V)_{iJ}\left( \overline{\tilde{\nu}}^i\gamma^{\mu}P_L\tilde{N}^{cJ} \right), 
\end{equation}
where $O_L$ is the matrix used to diagonalize the charged lepton sector: $O_L^{\dagger} m_l O_R = diag( m_e, m_{\mu}, m_{\tau} )$. 
These interactions lead to decay modes of heavy right-handed neutrinos governed by the couplings $(V^{\dagger}O_L)_{Ji}$. In the limit where the yukawa matrix $(m_d)_{iJ} \ll M_{IJ}$ the approximate form of V is given by $V_{iJ} = (m_d)_{iI}(M^{-1})_{IJ}$. In order to streamline our discussion in the text, we shall define ${\bf V} \equiv \big[ (m_d)(M^{-1})\big]^{\dagger}O_L$ and $ {\bf \tilde{V} } \equiv U^{\dagger} (m_d) (M^{-1})$. In this case the interactions read: 
\begin{equation}
\mathcal{L} \supset \frac{g}{\sqrt{2}}W_{\mu}^+({\bf V})_{Ji}\left( \overline{\tilde{N^c}}^{J}\gamma^{\mu}P_L\tilde{l}^i \right)  +\frac{g}{2\cos{\theta_W}}Z_{\mu}({\bf \tilde{V}})_{iJ}\left( \overline{\tilde{\nu}}^i\gamma^{\mu}P_L\tilde{N}^{cJ} \right). 
\end{equation}

%\begin{fmffile}{feyngraph}
 % \begin{fmfgraph}(100,100)
   %      \fmfleft{i}
     %    \fmfright{o0,o1,o2,o3,o4,o5}
     %    \fmf{dashes,tension=4}{i,v1}
       %  \fmf{fermion,tension=0.5}{o5,v3}
         %\fmf{fermion,tension=0.5}{v2,o0}
  %       \fmf{plain,tension=2}{v1,v2}
    %     \fmf{plain,tension=2}{v1,v3}
      %   \fmf{photon}{v2,v6}   
%         \fmf{photon}{v3,v7}      
%         \fmf{fermion,tension=0.3}{o2,v6,o1}
%         \fmf{fermion,tension=0.3}{o3,v7,o4}
%  \end{fmfgraph}
% \end{fmffile}

%\begin{fmffile}{feyngraph}
  %\begin{fmfgraph*}(200,200)
%         \fmfleft{i}
 %        \fmfright{o0,o1,o2,o3,o4,o5}
  %       \fmfbottom{b0,b1,b2,b3,b4,b5}
  %       \fmftop{t0,t1,t2,t3,t4,t5}
  %       \fmf{dashes,tension=1,label=$h$}{i,v1}
    %     \fmf{fermion,tension=0.5,label=$e^+$}{t4,v3}
 %%        \fmf{fermion,tension=0.5,label=$e^-$}{v2,b4}
 %        \fmf{plain,tension=1,label=$N_1$,label.dist=1}{v1,v2}
   %      \fmf{plain,tension=1,label=$N_1$,label.side=left,label.dist=1}{v1,v3}
   %%      \fmf{photon,label=$W^{*+}$,label.side=left}{v2,v6}   
    %     \fmf{photon,label=$W^{*-}$,label.side=right}{v3,v7}      
     %    \fmf{fermion,tension=0.7}{o2,v6,o1}
     %    \fmf{fermion,tension=0.7}{o3,v7,o4}
    %     \fmflabel{$e^-$}{o4} 
      %   \fmflabel{$\overline{\nu_e}$}{o3}
    %     \fmflabel{$u$}{o2} 
    %     \fmflabel{$\overline{d}$}{o1} 
           
 % \end{fmfgraph*}
%\end{fmffile}

\bibliography{references}

\providecommand{\href}[2]{#2}\begingroup\raggedright\begin{thebibliography}{10}

\bibitem{Minkowski:1977sc}
P.~Minkowski, \emph{{$\mu \to e\gamma$ at a Rate of One Out of $10^{9}$ Muon
  Decays?}}, \href{https://doi.org/10.1016/0370-2693(77)90435-X}{\emph{Phys.
  Lett.} {\bfseries 67B} (1977) 421}.

\bibitem{Mohapatra:1979ia}
R.~N. Mohapatra and G.~Senjanovic, \emph{{Neutrino Mass and Spontaneous Parity
  Nonconservation}},
  \href{https://doi.org/10.1103/PhysRevLett.44.912}{\emph{Phys. Rev. Lett.}
  {\bfseries 44} (1980) 912}.

\bibitem{Tanabashi:2018oca}
{\scshape Particle Data Group} collaboration, \emph{{Review of Particle
  Physics}}, \href{https://doi.org/10.1103/PhysRevD.98.030001}{\emph{Phys.
  Rev.} {\bfseries D98} (2018) 030001}.

\bibitem{Mohapatra:2006gs}
R.~N. Mohapatra and A.~Y. Smirnov, \emph{{Neutrino Mass and New Physics}},
  \href{https://doi.org/10.1146/annurev.nucl.56.080805.140534}{\emph{Ann. Rev.
  Nucl. Part. Sci.} {\bfseries 56} (2006) 569}
  [\href{https://arxiv.org/abs/hep-ph/0603118}{{\ttfamily hep-ph/0603118}}].

\bibitem{Pilaftsis:1991ug}
A.~Pilaftsis, \emph{{Radiatively induced neutrino masses and large Higgs
  neutrino couplings in the standard model with Majorana fields}},
  \href{https://doi.org/10.1007/BF01482590}{\emph{Z. Phys.} {\bfseries C55}
  (1992) 275} [\href{https://arxiv.org/abs/hep-ph/9901206}{{\ttfamily
  hep-ph/9901206}}].

\bibitem{Belotsky:2002ym}
K.~Belotsky, D.~Fargion, M.~Khlopov, R.~Konoplich and K.~Shibaev,
  \emph{{Invisible Higgs boson decay into massive neutrinos of fourth
  generation}}, \href{https://doi.org/10.1103/PhysRevD.68.054027}{\emph{Phys.
  Rev.} {\bfseries D68} (2003) 054027}
  [\href{https://arxiv.org/abs/hep-ph/0210153}{{\ttfamily hep-ph/0210153}}].

\bibitem{Han:2006ip}
T.~Han and B.~Zhang, \emph{{Signatures for Majorana neutrinos at hadron
  colliders}}, \href{https://doi.org/10.1103/PhysRevLett.97.171804}{\emph{Phys.
  Rev. Lett.} {\bfseries 97} (2006) 171804}
  [\href{https://arxiv.org/abs/hep-ph/0604064}{{\ttfamily hep-ph/0604064}}].

\bibitem{Hung:2006ap}
P.~Q. Hung, \emph{{A Model of electroweak-scale right-handed neutrino mass}},
  \href{https://doi.org/10.1016/j.physletb.2007.03.067}{\emph{Phys. Lett.}
  {\bfseries B649} (2007) 275}
  [\href{https://arxiv.org/abs/hep-ph/0612004}{{\ttfamily hep-ph/0612004}}].

\bibitem{Graesser:2007yj}
M.~L. Graesser, \emph{{Broadening the Higgs boson with right-handed neutrinos
  and a higher dimension operator at the electroweak scale}},
  \href{https://doi.org/10.1103/PhysRevD.76.075006}{\emph{Phys. Rev.}
  {\bfseries D76} (2007) 075006}
  [\href{https://arxiv.org/abs/0704.0438}{{\ttfamily 0704.0438}}].

\bibitem{Graesser:2007pc}
M.~L. Graesser, \emph{{Experimental Constraints on Higgs Boson Decays to
  TeV-scale Right-Handed Neutrinos}},
  \href{https://arxiv.org/abs/0705.2190}{{\ttfamily 0705.2190}}.

\bibitem{Helo:2013esa}
J.~C. Helo, M.~Hirsch and S.~Kovalenko, \emph{{Heavy neutrino searches at the
  LHC with displaced vertices}},
  \href{https://doi.org/10.1103/PhysRevD.89.073005,
  10.1103/PhysRevD.93.099902}{\emph{Phys. Rev.} {\bfseries D89} (2014) 073005}
  [\href{https://arxiv.org/abs/1312.2900}{{\ttfamily 1312.2900}}].

\bibitem{Maiezza:2015lza}
A.~Maiezza, M.~Nemevšek and F.~Nesti, \emph{{Lepton Number Violation in Higgs
  Decay at LHC}},
  \href{https://doi.org/10.1103/PhysRevLett.115.081802}{\emph{Phys. Rev. Lett.}
  {\bfseries 115} (2015) 081802}
  [\href{https://arxiv.org/abs/1503.06834}{{\ttfamily 1503.06834}}].

\bibitem{Accomando:2016rpc}
E.~Accomando, L.~Delle~Rose, S.~Moretti, E.~Olaiya and C.~H.
  Shepherd-Themistocleous, \emph{{Novel SM-like Higgs decay into displaced
  heavy neutrino pairs in U(1) models}},
  \href{https://doi.org/10.1007/JHEP04(2017)081}{\emph{JHEP} {\bfseries 04}
  (2017) 081} [\href{https://arxiv.org/abs/1612.05977}{{\ttfamily
  1612.05977}}].

\bibitem{Caputo:2017pit}
A.~Caputo, P.~Hernandez, J.~Lopez-Pavon and J.~Salvado, \emph{{The seesaw
  portal in testable models of neutrino masses}},
  \href{https://doi.org/10.1007/JHEP06(2017)112}{\emph{JHEP} {\bfseries 06}
  (2017) 112} [\href{https://arxiv.org/abs/1704.08721}{{\ttfamily
  1704.08721}}].

\bibitem{Deppisch:2018eth}
F.~F. Deppisch, W.~Liu and M.~Mitra, \emph{{Long-lived Heavy Neutrinos from
  Higgs Decays}}, \href{https://doi.org/10.1007/JHEP08(2018)181}{\emph{JHEP}
  {\bfseries 08} (2018) 181}
  [\href{https://arxiv.org/abs/1804.04075}{{\ttfamily 1804.04075}}].

\bibitem{Cottin:2018kmq}
G.~Cottin, J.~C. Helo and M.~Hirsch, \emph{{Searches for light sterile
  neutrinos with multitrack displaced vertices}},
  \href{https://doi.org/10.1103/PhysRevD.97.055025}{\emph{Phys. Rev.}
  {\bfseries D97} (2018) 055025}
  [\href{https://arxiv.org/abs/1801.02734}{{\ttfamily 1801.02734}}].

\bibitem{Cottin:2018nms}
G.~Cottin, J.~C. Helo and M.~Hirsch, \emph{{Displaced vertices as probes of
  sterile neutrino mixing at the LHC}},
  \href{https://doi.org/10.1103/PhysRevD.98.035012}{\emph{Phys. Rev.}
  {\bfseries D98} (2018) 035012}
  [\href{https://arxiv.org/abs/1806.05191}{{\ttfamily 1806.05191}}].

\bibitem{Drewes:2019fou}
M.~Drewes and J.~Hajer, \emph{{Heavy Neutrinos in displaced vertex searches at
  the LHC and HL-LHC}},  \href{https://arxiv.org/abs/1903.06100}{{\ttfamily
  1903.06100}}.

\bibitem{Liu:2019ayx}
J.~Liu, Z.~Liu, L.-T. Wang and X.-P. Wang, \emph{{Seeking for sterile neutrinos
  with displaced leptons at the LHC}},
  \href{https://arxiv.org/abs/1904.01020}{{\ttfamily 1904.01020}}.

\bibitem{Dine:1994vc}
M.~Dine, A.~E. Nelson and Y.~Shirman, \emph{{Low-energy dynamical supersymmetry
  breaking simplified}},
  \href{https://doi.org/10.1103/PhysRevD.51.1362}{\emph{Phys. Rev.} {\bfseries
  D51} (1995) 1362} [\href{https://arxiv.org/abs/hep-ph/9408384}{{\ttfamily
  hep-ph/9408384}}].

\bibitem{Toback:2004xd}
D.~A. Toback and P.~Wagner, \emph{{Prospects of searches for neutral,
  long-lived particles which decay to photons using timing at CDF}},
  \href{https://doi.org/10.1103/PhysRevD.70.114032}{\emph{Phys. Rev.}
  {\bfseries D70} (2004) 114032}
  [\href{https://arxiv.org/abs/hep-ph/0407022}{{\ttfamily hep-ph/0407022}}].

\bibitem{Abulencia:2007ut}
{\scshape CDF} collaboration, \emph{{Search for heavy, long-lived particles
  that decay to photons at CDF II}},
  \href{https://doi.org/10.1103/PhysRevLett.99.121801}{\emph{Phys. Rev. Lett.}
  {\bfseries 99} (2007) 121801}
  [\href{https://arxiv.org/abs/0704.0760}{{\ttfamily 0704.0760}}].

\bibitem{Aaltonen:2008dm}
{\scshape CDF} collaboration, \emph{{Search for Heavy, Long-Lived Neutralinos
  that Decay to Photons at CDF II Using Photon Timing}},
  \href{https://doi.org/10.1103/PhysRevD.78.032015}{\emph{Phys. Rev.}
  {\bfseries D78} (2008) 032015}
  [\href{https://arxiv.org/abs/0804.1043}{{\ttfamily 0804.1043}}].

\bibitem{CMS-PAS-EXO-12-035}
{\scshape CMS Collaboration} collaboration, \emph{{Search for long-lived
  neutral particles in the final state of delayed photons and missing energy in
  proton-proton collisions at $\sqrt s$ = 8 TeV}},  Tech. Rep.
  CMS-PAS-EXO-12-035, CERN, Geneva, 2015.

\bibitem{cms_mip}
C.~Collaboration, \emph{{TECHNICAL PROPOSAL FOR A MIP TIMING DETECTOR IN THE
  CMS EXPERIMENT PHASE 2 UPGRADE}},  Tech. Rep. CERN-LHCC-2017-027. LHCC-P-009,
  CERN, Geneva, Dec, 2017.

\bibitem{Liu:2018wte}
J.~Liu, Z.~Liu and L.-T. Wang, \emph{{Long-lived particles at the LHC: catching
  them in time}},
  \href{https://doi.org/10.1103/PhysRevLett.122.131801}{\emph{Phys. Rev. Lett.}
  {\bfseries 122} (2019) 131801}
  [\href{https://arxiv.org/abs/1805.05957}{{\ttfamily 1805.05957}}].

\bibitem{Chatrchyan:2014fea}
{\scshape CMS} collaboration, \emph{{Description and performance of track and
  primary-vertex reconstruction with the CMS tracker}},
  \href{https://doi.org/10.1088/1748-0221/9/10/P10009}{\emph{JINST} {\bfseries
  9} (2014) P10009} [\href{https://arxiv.org/abs/1405.6569}{{\ttfamily
  1405.6569}}].

\bibitem{Alimena:2019zri}
J.~Alimena et~al., \emph{{Searching for long-lived particles beyond the
  Standard Model at the Large Hadron Collider}},
  \href{https://arxiv.org/abs/1903.04497}{{\ttfamily 1903.04497}}.

\bibitem{Aad:2015gba}
{\scshape ATLAS} collaboration, \emph{{Measurements of the Higgs boson
  production and decay rates and coupling strengths using pp collision data at
  $\sqrt{s}=7$ and 8 TeV in the ATLAS experiment}},
  \href{https://doi.org/10.1140/epjc/s10052-015-3769-y}{\emph{Eur. Phys. J.}
  {\bfseries C76} (2016) 6} [\href{https://arxiv.org/abs/1507.04548}{{\ttfamily
  1507.04548}}].

\bibitem{Duarte:2016miz}
L.~Duarte, I.~Romero, J.~Peressutti and O.~A. Sampayo, \emph{{Effective
  Majorana neutrino decay}},
  \href{https://doi.org/10.1140/epjc/s10052-016-4301-8}{\emph{Eur. Phys. J.}
  {\bfseries C76} (2016) 453}
  [\href{https://arxiv.org/abs/1603.08052}{{\ttfamily 1603.08052}}].

\bibitem{Duarte:2015iba}
L.~Duarte, J.~Peressutti and O.~A. Sampayo, \emph{{Majorana neutrino decay in
  an Effective Approach}},
  \href{https://doi.org/10.1103/PhysRevD.92.093002}{\emph{Phys. Rev.}
  {\bfseries D92} (2015) 093002}
  [\href{https://arxiv.org/abs/1508.01588}{{\ttfamily 1508.01588}}].

\bibitem{atlas_trigger}
A.~Collaboration, \emph{Trigger menu in 2016}, .

\bibitem{Khachatryan:2016bia}
{\scshape CMS} collaboration, \emph{{The CMS trigger system}},
  \href{https://doi.org/10.1088/1748-0221/12/01/P01020}{\emph{JINST} {\bfseries
  12} (2017) P01020} [\href{https://arxiv.org/abs/1609.02366}{{\ttfamily
  1609.02366}}].

\bibitem{CMS:2014hka}
{\scshape CMS} collaboration, \emph{{Search for long-lived particles that decay
  into final states containing two electrons or two muons in proton-proton
  collisions at $\sqrt{s} =$ 8 TeV}},
  \href{https://doi.org/10.1103/PhysRevD.91.052012}{\emph{Phys. Rev.}
  {\bfseries D91} (2015) 052012}
  [\href{https://arxiv.org/abs/1411.6977}{{\ttfamily 1411.6977}}].

\bibitem{Khachatryan:2014mea}
{\scshape CMS} collaboration, \emph{{Search for Displaced Supersymmetry in
  events with an electron and a muon with large impact parameters}},
  \href{https://doi.org/10.1103/PhysRevLett.114.061801}{\emph{Phys. Rev. Lett.}
  {\bfseries 114} (2015) 061801}
  [\href{https://arxiv.org/abs/1409.4789}{{\ttfamily 1409.4789}}].

\bibitem{Alloul:2013bka}
A.~Alloul, N.~D. Christensen, C.~Degrande, C.~Duhr and B.~Fuks,
  \emph{{FeynRules 2.0 - A complete toolbox for tree-level phenomenology}},
  \href{https://doi.org/10.1016/j.cpc.2014.04.012}{\emph{Comput. Phys. Commun.}
  {\bfseries 185} (2014) 2250}
  [\href{https://arxiv.org/abs/1310.1921}{{\ttfamily 1310.1921}}].

\bibitem{Alva:2014gxa}
D.~Alva, T.~Han and R.~Ruiz, \emph{{Heavy Majorana neutrinos from $W\gamma$
  fusion at hadron colliders}},
  \href{https://doi.org/10.1007/JHEP02(2015)072}{\emph{JHEP} {\bfseries 02}
  (2015) 072} [\href{https://arxiv.org/abs/1411.7305}{{\ttfamily 1411.7305}}].

\bibitem{Degrande:2016aje}
C.~Degrande, O.~Mattelaer, R.~Ruiz and J.~Turner, \emph{{Fully-Automated
  Precision Predictions for Heavy Neutrino Production Mechanisms at Hadron
  Colliders}}, \href{https://doi.org/10.1103/PhysRevD.94.053002}{\emph{Phys.
  Rev.} {\bfseries D94} (2016) 053002}
  [\href{https://arxiv.org/abs/1602.06957}{{\ttfamily 1602.06957}}].

\bibitem{Atre:2009rg}
A.~Atre, T.~Han, S.~Pascoli and B.~Zhang, \emph{{The Search for Heavy Majorana
  Neutrinos}}, \href{https://doi.org/10.1088/1126-6708/2009/05/030}{\emph{JHEP}
  {\bfseries 05} (2009) 030} [\href{https://arxiv.org/abs/0901.3589}{{\ttfamily
  0901.3589}}].

\bibitem{Alwall:2014hca}
J.~Alwall, R.~Frederix, S.~Frixione, V.~Hirschi, F.~Maltoni, O.~Mattelaer
  et~al., \emph{{The automated computation of tree-level and next-to-leading
  order differential cross sections, and their matching to parton shower
  simulations}}, \href{https://doi.org/10.1007/JHEP07(2014)079}{\emph{JHEP}
  {\bfseries 07} (2014) 079} [\href{https://arxiv.org/abs/1405.0301}{{\ttfamily
  1405.0301}}].

\bibitem{Sjostrand:2014zea}
T.~Sjöstrand, S.~Ask, J.~R. Christiansen, R.~Corke, N.~Desai, P.~Ilten et~al.,
  \emph{{An Introduction to PYTHIA 8.2}},
  \href{https://doi.org/10.1016/j.cpc.2015.01.024}{\emph{Comput. Phys. Commun.}
  {\bfseries 191} (2015) 159}
  [\href{https://arxiv.org/abs/1410.3012}{{\ttfamily 1410.3012}}].

\bibitem{deFavereau:2013fsa}
{\scshape DELPHES 3} collaboration, \emph{{DELPHES 3, A modular framework for
  fast simulation of a generic collider experiment}},
  \href{https://doi.org/10.1007/JHEP02(2014)057}{\emph{JHEP} {\bfseries 02}
  (2014) 057} [\href{https://arxiv.org/abs/1307.6346}{{\ttfamily 1307.6346}}].

\bibitem{deFlorian:2016spz}
{\scshape LHC Higgs Cross Section Working Group} collaboration, \emph{{Handbook
  of LHC Higgs Cross Sections: 4. Deciphering the Nature of the Higgs Sector}},
   \href{https://arxiv.org/abs/1610.07922}{{\ttfamily 1610.07922}}.

\bibitem{Khachatryan:2016mlc}
{\scshape CMS} collaboration, \emph{{Measurement and QCD analysis of
  double-differential inclusive jet cross sections in pp collisions at $
  \sqrt{s}=8 $ TeV and cross section ratios to 2.76 and 7 TeV}},
  \href{https://doi.org/10.1007/JHEP03(2017)156}{\emph{JHEP} {\bfseries 03}
  (2017) 156} [\href{https://arxiv.org/abs/1609.05331}{{\ttfamily
  1609.05331}}].

\bibitem{Aad:2016xcr}
{\scshape ATLAS} collaboration, \emph{{Measurement of the inclusive isolated
  prompt photon cross section in pp collisions at $ \sqrt{s}=8 $ TeV with the
  ATLAS detector}}, \href{https://doi.org/10.1007/JHEP08(2016)005}{\emph{JHEP}
  {\bfseries 08} (2016) 005}
  [\href{https://arxiv.org/abs/1605.03495}{{\ttfamily 1605.03495}}].

\bibitem{jetfake}
C.~Jessup, ``{Measurement of the inclusive isolated prompt photon cross section
  in pp collisions at $ \sqrt{s}=8 $ TeV with the ATLAS detector}.''
  {https://www3.nd.edu/~cjessop/talks/JessopCornell.pdf}, 2007.

\bibitem{CMS-PAS-EXO-19-001}
{\scshape CMS Collaboration} collaboration, \emph{{Search for long-lived
  particles using delayed jets and missing transverse momentum with
  proton-proton collisions at $\sqrt{s}$ = 13 TeV}},  Tech. Rep.
  CMS-PAS-EXO-19-001, CERN, Geneva, 2019.

\bibitem{Aghanim:2018eyx}
{\scshape Planck} collaboration, \emph{{Planck 2018 results. VI. Cosmological
  parameters}},  \href{https://arxiv.org/abs/1807.06209}{{\ttfamily
  1807.06209}}.

\bibitem{Curtin:2018mvb}
D.~Curtin et~al., \emph{{Long-Lived Particles at the Energy Frontier: The
  MATHUSLA Physics Case}},  \href{https://arxiv.org/abs/1806.07396}{{\ttfamily
  1806.07396}}.

\end{thebibliography}\endgroup
\bibliographystyle{jhep}
\end{document}